\documentclass[a4paper,12pt]{article}
\usepackage{amssymb}
\usepackage{latexsym}

\usepackage{epsfig}
\usepackage{graphicx}

\usepackage{amsmath,amsthm,amsfonts,amssymb}

\newcommand{\be}{\begin{eqnarray}}
\newcommand{\ee}{\end{eqnarray}}
\newcommand{\bea}{\begin{eqnarray}}
\newcommand{\nn}{\nonumber}
\newcommand{\eea}{\end{eqnarray}}

\newcommand{\nk}{\noindent}

\begin{document}

\begin{titlepage}
\begin{flushright}
hep-th/0208207\\ August 2002
\end{flushright}
\begin{centering}
\vspace{.8in}
{\large {\bf Black Hole Solutions in Braneworlds with Induced
Gravity}}
\\

\vspace{.5in} {\bf G. Kofinas\footnote{kofinas@physics.uoc.gr},
E. Papantonopoulos\footnote{lpapa@central.ntua.gr},  V.
Zamarias\footnote{zamarias@central.ntua.gr}}

\vspace{0.3in}
$^{1}$ Department of Physics and Institute of Plasma Physics,\\
University of Crete, 710 03 Heraklion, Greece\\
$^{2,\,3}$ Department of Physics, National Technical University of Athens,\\
Zografou Campus GR 157 80, Athens, Greece\\
\end{centering}

\vspace{1in}
%%%%%%%%%%%%%%%%%%%%%%% ABSTRACT %%%%%%%%%%%%%%%%%%%%%%%%%%%%%%%%%%%
\begin{abstract}
\nk We extent our previous study on spherically symmetric
braneworld solutions with induced gravity, including non-local
bulk effects. We find the most general static four-dimensional
black hole solutions with $g_{tt}=-g_{rr}^{-1}$. They satisfy a
closed system of equations on the brane and represent the
strong-gravity corrections to the Schwarzschild-$(A)dS_{4}$
spacetime. These new solutions have extra terms which give extra
attraction relative to the Newtonian-$(A)dS_{4}$ force; however,
the conventional limits are easily obtained. These terms, when
defined asymptotically, behave like $AdS_{4}$ in this regime,
while when defined at infinitely short distances predict either an
additional attractive Newtonian potential or an attractive
potential which scales approximately as $\sqrt{r}$. One of the
solutions found gives extra deflection of light compared to
Newtonian deflection.
\end{abstract}
%%%%%%%%%%%%%%%%%%%%%%%%%%%%%%%%%%%%%%%%%%%%%%%%%%%%%%%%%%%%%%%%%%%%%
\end{titlepage}

\newpage

\baselineskip=18pt

%%%%%%%%%%%%%%%%%%INTRODUCTION %%%%%%%%%%%%%%%%%%%%%%%%%%%%%%%%%%

\section*{1. \,\,\,Introduction}
 \hspace{0.8cm}
Branes are solitonic solutions of ten-dimensional string theories.
In the most simplified picture of the braneworld scenario, our
physical world is realized as a four-dimensional hypersurface
embedded in a five-dimensional space called bulk. All matter and
gauge interactions live on the brane, while the gravitational
interactions are effective in the whole five-dimensional space.
This novel approach of visualizing our world, offers a new
understanding of the four fundamental forces. While in the
so-called symmetric picture all known interactions were tried to
be unified under the same symmetry group, the braneworld scenario
treats the weak, electromagnetic and strong interactions
differently than gravitational interactions. This allowed to
define a gravitational scale of the whole space, which if the
extra dimensions are large, can be as low as the $TeV$ scale
\cite{dimo, randall, horava}, while the four-dimensional
gravitational scale of our world is at the Planck scale. This
happens because our four-dimensional world is confined on the
brane and can ``see'' only the four-dimensional localized
gravitational field.
\par
Braneworld solutions can give us information about the structure
and nature of the extra dimensions. Very recently, we have a
plethora of observational data, both cosmological and
astrophysical. Consistency of cosmological and local
braneworld solutions with these data can give information about
the parameters of the theory, like the energy scale, the size of
the extra dimensions or the strength of the gravitational force of
the extra dimensions. For example, spherically symmetric local
braneworld solutions can give information about the crossover
scale above which the extra dimensions appear, how the Newton's
constant changes with matter density, or what are the corrections
of the gravitational potential at high energies.
\par
Braneworld solutions can be obtained following two different
approaches. In the first approach, the dynamics and the geometry
of the whole space is primarily considered, and then, the dynamics
on the brane is extracted using mainly consistency checks like the
Israel matching conditions. The second approach is to specify the
dynamics and the geometry on the brane first, and then try to
extent the solution to the bulk. A disadvantage of this method is
that finding the bulk geometry in which the brane consists its
boundary may be a very difficult task. Another difficulty in this
approach is, as we will discuss later, that it is not always
possible to obtain a closed set of equations on the brane, so that
only with data on the brane to be able to predict the behavior of
the fields on the brane. However, this method is basically the
only way we have for finding non-trivial braneworld solutions (i.e.
solutions not arising from factorizable bulk geometries).
\par

The effective brane equations have been obtained \cite{maeda} when
the
effective low-energy theory in the bulk is higher-dimensional
gravity.
However, a more fundamental description of the physics
 that produces the
brane could include \cite{sundrum} higher order
 terms in a derivative
expansion of the effective action, such as a
term for the scalar curvature of
the brane, and higher powers of
curvature tensors on the brane. If the
dynamics is governed not
 only by the ordinary five-dimensional
Einstein-Hilbert action, but
 also by the four-dimensional Ricci scalar term
induced on the
 brane, new phenomena appear. In \cite{dvali1}, it was
observed that the localized matter fields on the brane (which
couple to bulk
gravitons) can generate via quantum loops a
localized four-dimensional
worldvolume kinetic term for gravitons
(see also \cite{capper, adler, zee,
khuri}). That is to say, four-dimensional gravity is induced from the bulk
gravity to the brane worldvolume
 by the matter fields confined to the brane.
It was also shown that
 an observer on the brane will see correct Newtonian
gravity at
 distances shorter than a certain crossover scale, despite the
fact
that gravity propagates in extra space which was assumed there to
 be
flat with infinite extent; at larger distances, the force
 becomes
higher-dimensional.
\par
A realization of the induced gravity scenario in string theory was
presented in \cite{kiritsis}. Furthermore, new closed string
couplings on
 Dp-branes for the bosonic string were found in
\cite{corley}.
These couplings are quadratic in derivatives and therefore
take
 the form of induced kinetic terms on the brane. For the graviton in
particular these are the induced Einstein-Hilbert term as well as terms
quadratic in the second fundamental tensor. Considering
 the intrinsic
curvature scalar in the bulk action, the effective
 brane equations have been
obtained in \cite{kofinas}. Results
 concerning cosmology have been discussed
in \cite{collins, shtanov, nojiri, deffayet, myung, dedg, def}.
\par
In our previous paper \cite{pappa}, we discussed the gravitational
field of an uncharged, non-rotating spherically symmetric rigid
object when in the dynamics there is a contribution from the brane
intrinsic curvature invariant. We found all the possible exterior
braneworld solutions. Some of these solutions are of the
Schwarzschild-$(A)dS_{4}$ form. In two cases, we also solved the
interior problem which reduces to a generalization of the
Oppenheimer-Volkoff solution. It was shown that the gravitational
constant get corrected for very small matter densities. The
conventional solar system bounds of General Relativity have set
the crossover scale below the $TeV$ scale. All the above results
were obtained by setting the electric part of the Weyl tensor,
$\textsf{E}_{\,\mu \nu}$, vanishing on the brane, as the boundary
condition of the propagation equations in the bulk space.
\par
In the present paper, we generalize our study of spherically
symmetric braneworld solutions with induced gravity, by including
the non-local bulk effects, as they are expressed by a
non-vanishing electric part of the Weyl tensor on the brane. By
choosing $g_{tt}=-g_{rr}^{-1}$, the system of equations on the
brane becomes closed and all the possible static black hole
solutions are found for these metrics. These solutions have
generic new terms which give extra attractive force compared to
the Newtonian -$(A)dS_{4}$ force, and represent the strong-gravity
corrections to the Schwarzschild-$(A)dS_{4}$ spacetime. However,
the conventional limits are easily obtained. The new terms, when
defined asymptotically, behave like $AdS_{4}$ in this regime,
while when defined at infinitely short distances predict either an
additional attractive Newtonian potential or an attractive
potential which scales approximately as $\sqrt{r}$. One of the
solutions found gives extra deflection of light compared to
Newtonian deflection.
\par
The paper is organized as follows. In section 2, we give a general
introduction to the induced gravity formalism and we review some
of the solutions found in \cite{pappa}. In section 3, we include
the electric part of the Weyl tensor in spherically symmetric
braneworld configurations and we show how the modified Einstein
equations supplemented by the Bianchi identities constitute a
closed system of equations on the brane. In section 4, we present
our solutions and discuss their physical implications. Finally, in
the last section we summarize our results.

%%%%%%%%%%%%%%%%%%%%%% SPHERICAL SOLUTIONS %%%%%%%%%%%%%%%%%%%%%%
\section*{2. \,\,Induced Gravity Equations on the Brane}
\hspace{0.8cm} We consider a 3-dimensional brane $\Sigma$ embedded
in a 5-dimensional spacetime $M$. Capital Latin letters
$A,B,...=0,1,...,4$ will denote full spacetime, lower Greek
$\mu,\nu,...=0,1,...,3$ run over brane worldvolume, while lower
Latin ones span some 3-dimensional spacelike surfaces foliating
the brane, i.e. $i,j,...=1,...,3$. For convenience, we can quite
generally, choose a coordinate $y$ such that the hypersurface
$y=0$ coincides with the brane. The total action for the system is
taken to be: \be
S&=&\frac{1}{2\kappa_{5}^{2}}\int_{M}\sqrt{-^{(5)}g}\,\,(^{(5)}R-2\Lambda_{5})d^{5}x+
\frac{1}{2\kappa_{4}^{2}}\int_{\Sigma}\sqrt{-^{(4)}g}\,\,(^{(4)}R-2\Lambda_{4})d^{4}x\nn\\&&+
\int_{M}\sqrt{-^{(5)}g}\,\,L_{5}^{mat}\,d^{5}x+
\int_{\Sigma}\sqrt{-^{(4)}g}\,\,L_{4}^{mat}\,d^{4}x.
\label{action} \ee For clarity, we have separated the cosmological
constants $\Lambda_{5}$, $\Lambda_{4}$ from the rest matter
contents $L_{5}^{mat}$, $L_{4}^{mat}$ of the bulk and the brane
respectively. $\Lambda_{4}/\kappa_{4}^{2}$ can be interpreted as
the brane tension of the standard Dirac-Nambu-Goto action and can
include quantum contributions to the four-dimensional cosmological
constant. We basically concern on the case with no fields in the
bulk, i.e. $^{(5)}T_{AB}=0$. \par From the dimensionful constants
$\kappa_{5}^{2}$, $\kappa_{4}^{2}$ the Planck masses $M_{5}$,
$M_{4}$ are defined as:
 \be
  \kappa_{5}^{2}=8\pi
G_{(5)}=M_{5}^{-3}\,\,\,\,\,,\,\,\,\,\, \kappa_{4}^{2}=8\pi
G_{(4)}=M_{4}^{-2}, \label{planck} \ee with $M_{5}$, $M_{4}$
having dimensions of (length)$^{-1}$. Then, a distance scale
$r_{c}$ is defined as :
 \be
r_{c}\equiv\frac{\kappa_{5}^{2}}{\kappa_{4}^{2}}=\frac{M_{4}^{2}}
{M_{5}^{3}}\,.
 \label{distancescale}
 \ee
Varying (\ref{action}) with respect to the bulk metric $g_{AB}$,
we obtain
the equations

\be
^{(5)}G_{AB}=-\Lambda_{5}g_{AB}+\kappa_{5}^{2}\,(^{(5)}T_{AB}+\,^{(loc)}T_{AB}\,\delta(y))\,,
\label{varying} \ee where \be
^{(loc)}T_{AB}\equiv-\frac{1}{\kappa_{4}^{2}}\,\sqrt{\frac{-^{(4)}g}
{- ^{(5)}g}}\,\,(^{(4)}G_{AB}-\kappa_{4}^{2}\,^{(4)}T_{AB}+
\Lambda_{4}h_{AB}) \label{tlocal} \ee is the localized
energy-momentum tensor of the brane. $^{(5)}G_{AB}$,
$^{(4)}G_{AB}$ denote the Einstein tensors
 constructed from the bulk and the
brane metrics respectively. Clearly, $^{(4)}G_{AB}$ acts as an additional
source term for the
 brane through $^{(loc)}T_{AB}$. The tensor
$h_{AB}=g_{AB}-n_{A}n_{B}$ is the induced metric on the
hypersurfaces $y=$ constant, with $n^{A}$ the normal vector on
these.
\par
The way the $y$-coordinate has been defined, allows us to write,
at least in
the neighborhood of the brane, the 5-line element in
 the block diagonal form
\be
 ds_{(5)}^{2}=-N^{2}dt^{2}+g_{ij}dx^{i}dx^{j}+dy^{2}\,,
\label{lineelement} \ee where $N,g_{ij}$ are generally functions
of $t,x^{i},y$. The distributional character of the brane matter
content makes necessary for the compatibility of the bulk
equations
 (\ref{varying}) the following modified
(due to
 $^{(4)}G^{\mu}_{\nu}$) Israel-Darmois-Lanczos-Sen conditions
\cite{israel} \be
[K_{\nu}^{\mu}]=-\kappa_{5}^{2}\left(^{(loc)}T_{\nu}^{\mu}-
\frac{^{(loc)}T}{3}\delta_{\nu}^{\mu}\right)\,, \label{israel} \ee
where the bracket means discontinuity of the extrinsic curvature
$K_{\mu\nu}=\partial_{y}g_{\mu\nu}/2$ across $y=0$. A
$\mathbf{Z}_{2}$ symmetry on reflection around the brane is
considered throughout.
\par
One can derive from equations (\ref{varying}), (\ref{israel}) the
induced brane gravitational dynamics \cite{kofinas}, which
consists of a four-dimensional Einstein gravity, coupled to a
well-defined modified matter content. More explicitly, one gets
\be
^{(4)}G_{\nu}^{\mu}=\kappa_{4}^{2}\,^{(4)}T_{\nu}^{\mu}-\Big(\Lambda_{4}
+\frac{3}{2}\alpha^{2}\Big)\,\delta_{\nu}^{\mu}+
\alpha\Big(L_{\nu}^{\mu}+\frac{L}{2}\,\delta_{\nu}^{\mu}\Big)\,,
\label{einstein} \ee where $\alpha\equiv 2/r_{c}$, while the
quantities $L^{\mu}_{\nu}$ are related to the matter content of
the theory through the equation \be
L_{\lambda}^{\mu}L_{\nu}^{\lambda}-\frac{L^{2}}{4}\,\delta_{\nu}^{\mu}
=\mathcal{T}_{\nu}^{\mu}-\frac{1}{4}(3\alpha^{2}+2\mathcal{T}_{\lambda}^{\lambda})\,\delta_{\nu}^{\mu}\,,
\label{lll} \ee and $L\equiv L^{\mu}_{\mu}$. The quantities
$\mathcal{T}^{\mu}_{\nu}$ are given by the expression \be
\mathcal{T}_{\nu}^{\mu}&=&\Big(\Lambda_{4}-\frac{1}{2}\,\Lambda_{5}\Big)\delta_{\nu}^{\mu}
-\kappa_{4}^{2}\,^{(4)}T_{\nu}^{\mu}+\nn\\&&
+\frac{2}{3}\,\kappa_{5}^{2}\,\Big(\,^{(5)}\overline{T}\,_{\nu}^{\mu}
+\Big(\,^{(5)}\overline{T}\,_{y}^{y}-\frac{^{(5)}\overline{T}}{4}\Big)\,\delta_{\nu}^{\mu}\Big)
-\overline{\textsf{E}}^{\,\mu}_{\,\nu}\,, \label{energy} \ee with
$^{(5)}\overline{T}=\,^{(5)}\overline{T}\,_{A}^{A}\,,\,
^{(5)}\overline{T}\,_{B}^{A}= g^{AC}\,^{(5)}\overline{T}\,_{CB}$.
Bars over $^{(5)}T^{A}_{B}$ and the electric part
$\textsf{E}^{\,^{\mu}}_{\,\nu}=C^{\mu}_{A \nu B}n^{A}n^{B}$ of the
5-dimensional Weyl tensor $C^{A}_{B C D}$ mean that the quantities
 are
evaluated at $y=0$. $\overline{\textsf{E}}^{\,\mu}_{\,\nu}$
 carries the
influence of non-local gravitational degrees of
 freedom in the bulk onto the
brane \cite{maeda} and makes the
 brane equations (\ref{einstein}) not to be,
in general, closed.
 This means that there are bulk degrees of freedom which
cannot be
 predicted from data available on the brane. One needs to solve the
field equations in the bulk in order to determine
$\textsf{E}^{\,^{\mu}}_{\,\nu}$ on the brane \cite{roys}.
\par
Due to the contracted Bianchi identities, the following differential
equations
 among $L^{\mu}_{\nu}$ arise from (\ref{einstein}) :
\be
L^{\mu}_{\nu ; \,\mu}+\frac{L_{;\, \nu}}{2}=0\,.
\label{bianchi}
\ee
\par
The spherically symmetric braneworld line-element is \be
ds_{(4)}^{2}=-B(r)dt^{2}+A(r)dr^{2}+r^{2}(d \theta^{2}+\sin ^{2}
\theta d \phi^{2}). \label{spherical} \ee The matter content of
the 3-universe is considered to be a localized spherically
symmetric untilted perfect fluid (e.g. a star) $^{(4)}T_{\mu
\nu}=(\rho+p)u_{\mu}u_{\nu}+pg_{\mu \nu}$ ($u^{\mu}$ stands for
the four-velocity of the fluid) with $\rho=p=0$ for $r>R$, plus
the cosmological constant $\Lambda_{4}$. The matter content of the
bulk is a cosmological constant $\Lambda_{5}$. In \cite{pappa}, we
considered the case $\overline{\textsf{E}}^{\,\mu}_{\,\nu}=0$ as
the boundary condition of the propagation equations in the bulk
space. This is somehow simplified from the viewpoint of geometric
complexity, but it was the first step for investigating the
characteristics carried by the brane curvature invariant on the
local brane dynamics we are interested in. All the solutions
outside a static localized matter distribution were found. One of
these is the Schwarzschild-$(A)dS_{4}$ metric which is matched to
a modified Oppenheimer-Volkoff interior. We will review this
solution and in the next section we will discuss the strong
gravity corrections resulting from the presence of the electric
part of the Weyl tensor on the brane. The other solutions have $A
B\neq 1$ and will not be discussed here. The exterior solution was
found to be
\begin{equation}
B_{>}(r)= \frac{1}{A_{>}(r)}=1-\frac{\gamma}{r}-\beta
r^{2}\,\,\,,\,\,\,r\geq R \,,
\label{generalA>}
\end{equation}
where $\gamma$ is an integration constant and \be
\beta=\frac{1}{3}\Lambda_{4}+\frac{1}{2}\alpha^{2}-
\frac{\alpha}{2\sqrt{3}}\sqrt{4\Lambda_{4}-2\Lambda_{5}+3\alpha^{2}}\,.
\label{beta1} \ee Considering a uniform distribution
$\rho(r)=\rho_{o}=\frac{3M}{4\pi R^{3}}$ for the object, the
interior solution was found, and its matching to the above
exterior specified the integration constant $\gamma$. The result
is \be
\frac{1}{A_{<}(r)}=1-(\beta+\frac{\gamma}{R^{3}})\,r^{2}\,\,\,,\,\,\,r
\leq R\,, \label{A<} \ee \be
B_{<}(r)=\frac{1-\frac{\gamma}{R}-\beta R^{2}}{\Big(1+\frac{4\pi
R^{3}}{3M}p(r)\Big)^{2}}\,\,\,,\,\,\,r\leq R\,, \label{B<} \ee \be
p(r)=-\rho_{o}\frac{\sqrt{1-(\beta+\frac{\gamma}{R^{3}})r^{2}}-
\sqrt{1-(\beta+\frac{\gamma}{R^{3}})R^{2}}}{\sqrt{1-(\beta+\frac{\gamma}{R^{3}})r^{2}}-
\omega \sqrt{1-(\beta+\frac{\gamma}{R^{3}})R^{2}}}\,,
\label{pressure} \ee where \be \gamma=\frac{\kappa_{4}^{2}M}{4
\pi} + \frac{\alpha R^3
}{2\sqrt{3}}\sqrt{4\Lambda_{4}-2\Lambda_{5}+3\alpha^{2}} -
\frac{\alpha
R^3}{2\sqrt{3}}\sqrt{4\Lambda_{4}-2\Lambda_{5}+3\alpha^{2}+\frac{3\kappa_{4}^{2}M}{\pi
R^{3}}}\,, \label{gamma1} \ee \be
\omega^{-1}=1-\frac{2}{\kappa_{4}^{2}\rho_{o}}\left(\beta+\frac{\gamma}{R^3}\right)
\Big(1-\frac{\sqrt{3}\alpha}{\sqrt{4\Lambda_{4}-2\Lambda_{5}+3\alpha^{2}+4\kappa_{4}^{2}\rho_{o}}}
\Big)^{-1}\,. \label{omega} \ee The parameters $\gamma$ and
$\beta$ of the Schwarzschild-$(A)dS_{4}$ exterior solution
(\ref{generalA>}) can be constrained by solar system experiments.
The bounds obtained fix the crossover scale below the $TeV$ range.
The $(A)dS_{4}$ term $\beta r^2$ in this black hole solution gives
at large distances additional force compared to the ordinary
Newtonian force. Finally, the $\gamma$ parameter in the $1/r$ term
modifies the Newton's gravitational constant which, as it is seen
from (\ref{gamma1}), for small matter densities deviates
significantly from its conventional value.

\section*{3. \,\,Black Holes in Induced Gravity with Non-Local Bulk Effects}
\hspace{0.8cm} So far we considered local corrections to the
Einstein equations on the brane. The presence of the electric part
$\overline{\textsf{E}}^{\,\mu}_{\,\nu}$ of the Weyl tensor  in
(\ref{energy}) indicates the 5D gravitational stresses, which are
known as massive KK modes of the graviton. For brane observers,
these stresses are non-local. Local density inhomogeneities on the
brane generate Weyl curvature in the bulk that ``back-reacts"
non-locally on the brane \cite{roys, maartens, phpapado, cham,
deruelle, casadio, dadhich}. Therefore, in general, this term
cannot be ignored.
\par
With respect to the privileged direction
 $u^{\mu}$ defined by the perfect fluid, the symmetric
and traceless tensor
 $\overline{\textsf{E}}_{\mu \nu}$ is uniquely and
irreducibly
 decomposed as follows
 \be
 \overline{\textsf{E}}_{\mu
\nu}=\mathcal{U}\Big(u_{\mu}u_{\nu}+\frac{1}{3}\textsf{h}_{\mu
\nu}\Big)+\mathcal{P}_{\mu \nu}+2\mathcal{Q}_{(\mu}u_{\nu)}\,,
\label{electric} \ee where $\textsf{h}_{\mu \nu}=g_{\mu
\nu}+u_{\mu}u_{\nu}$ is the projection operator normal to
$u^{\mu}$, while $\mathcal{P}_{\mu
\nu}u^{\nu}=\mathcal{Q}_{\mu}u^{\mu}=0$. $\mathcal{U}$ is the
non-local energy density, $\mathcal{P}_{\mu \nu}$ the non-local
anisotropic stress, and $\mathcal{Q}_{\mu}$ the non-local energy
flux on the brane. Static spherical symmetry implies \cite{roys}
that \be \mathcal{Q}_{\mu}=0\,\,\,\,,\,\,\,\,\mathcal{P}_{\mu
\nu}=\mathcal{P}(r)\Big(r_{\mu}r_{\nu}-\frac{1}{3}\textsf{h}_{\mu
\nu}\Big)\,, \label{sphericalrestriction} \ee where $r_{\mu}$ is
the unit radial vector. Thus, the non-vanishing components of the
electric part of the Weyl tensor are \be
\overline{\textsf{E}}^{\,0}_{\,0}=-\mathcal{U}\,\,,\,\,\overline{\textsf{E}}^{\,r}_{\,r}=
\frac{1}{3}(\mathcal{U}+2\mathcal{P})\,\,,\,\,\overline{\textsf{E}}^{\,\theta}_{\,\theta}=
\overline{\textsf{E}}^{\,\phi}_{\,\phi}=\frac{1}{3}(\mathcal{U}-\mathcal{P})\,.
\label{electriccomponents} \ee This means that for
$^{(5)}\overline{T}^{A}_{B}=0$ in (\ref{energy}), the matrix
$\mathcal{T}^{i}_{j}$ has two distinct eigenvalues, namely \be
\mathcal{T}^{r}_{r}=\Lambda_{4}-\frac{1}{2}\Lambda_{5}-\kappa_{4}^{2}p-\frac{1}{3}(\mathcal{U}+2\mathcal{P})\equiv
\tau_{r} \label{tirr} \ee \be
\mathcal{T}^{\theta}_{\theta}=\mathcal{T}^{\phi}_{\phi}=\Lambda_{4}-\frac{1}{2}\Lambda_{5}-\kappa_{4}^{2}p-\frac{1}{3}(\mathcal{U}-\mathcal{P})\equiv
\tau_{\theta}\equiv \tau_{\phi}\,. \label{tithth} \ee
\par
Consequently, due to the block-diagonal form of the metric, the
solution of the algebraic system (\ref{lll}) is \be
L^{0}_{i}=L^{i}_{0}=0\,, \label{l0i} \ee together with one of the
following three distinct cases for $L^{i}_{j}$, $L^{0}_{0}$\,:
\newline
\underline{Case 1} \be
L^{i}_{j}=diag\left(\sqrt{(L^{0}_{0})^{2}+\tau_{r}-\mathcal{T}^{0}_{0}}\,,\,
\sqrt{(L^{0}_{0})^{2}+\tau_{\theta}-\mathcal{T}^{0}_{0}}\,,\,
\sqrt{(L^{0}_{0})^{2}+\tau_{\theta}-\mathcal{T}^{0}_{0}}\right)
\label{case1} \ee \underline{Case 2} \be
L^{i}_{j}=diag\left(\sqrt{(L^{0}_{0})^{2}+\tau_{r}-\mathcal{T}^{0}_{0}}\,,\,
\sqrt{(L^{0}_{0})^{2}+\tau_{\theta}-\mathcal{T}^{0}_{0}}\,,\,
-\sqrt{(L^{0}_{0})^{2}+\tau_{\theta}-\mathcal{T}^{0}_{0}}\right)
\label{case2} \ee \underline{Case 3} \be
L^{i}_{j}=diag\left(\sqrt{(L^{0}_{0})^{2}+\tau_{r}-\mathcal{T}^{0}_{0}}\,,\,
-\sqrt{(L^{0}_{0})^{2}+\tau_{\theta}-\mathcal{T}^{0}_{0}}\,,\,
-\sqrt{(L^{0}_{0})^{2}+\tau_{\theta}-\mathcal{T}^{0}_{0}}\right)
\label{case3} \ee and the algebraic equation
\be
\sqrt{(L^{0}_{0})^{2}+\tau_{r}-\mathcal{T}^{0}_{0}}+2\epsilon
\sqrt{(L^{0}_{0})^{2}+\tau_{\theta}-\mathcal{T}^{0}_{0}}\pm
L^{0}_{0}= \sqrt{4(L^{0}_{0})^{2}+2(\tau_{r}+2\tau_{\theta})+3
\alpha^{2}-2\mathcal{T}^{0}_{0}} \label{algebraic} \ee with
$\epsilon = +1, 0, -1$ for the cases 1,2,3 respectively.
\par
For the considered metric (\ref{spherical}), one evaluates the
Ricci tensor $^{(4)}R_{\mu \nu}$, and construct the field
equations (\ref{einstein}). The combination
$^{(4)}R_{rr}/2A\,+\,^{(4)}R_{\theta\theta}/r^{2}\,+\,^{(4)}R_{00}/2B$
provides the following differential equation for $A(r)$: \be
\frac{A'}{A}=\frac{1-A}{r}+Ar\Big[\kappa_{4}^{2}
\rho+\Lambda_{4}+\frac{3}{2}\alpha^{2}-\frac{\alpha}{2}(3L^{0}_{0}+L^{r}_{r}+
2L^{\theta}_{\theta})\Big]\,, \label{A} \ee $\Big(\,'\equiv
\frac{d}{dr}\Big)$. Eliminating $\frac{A'}{A}$ in the $(\theta
\theta)$ component of (\ref{einstein}) using (\ref{A}), we get an
equation for $\frac{B'}{B}$, from which we obtain \be
\frac{(AB)\,'}{AB}=Ar\,[\,\kappa_{4}^{2}(\rho+p)-\alpha
(L^{0}_{0}-L^{r}_{r})\,]. \label{AB} \ee From the $(\theta
\theta)$, $(\phi \phi)$ equations of (\ref{einstein}) it arises
that \be L^{\theta}_{\theta}=L^{\phi}_{\phi}\,. \label{lthth} \ee
Using (\ref{lthth}), the Bianchi equations (\ref{bianchi}) take
the form \be
\frac{B'}{B}(L^{r}_{r}-L^{0}_{0})+\frac{4}{r}(L^{r}_{r}-L^{\theta}_{\theta})+(L^{0}_{0}+3L^{r}_{r}+2L^{\theta}_{\theta})\,'=0\,.
\label{bianchi1} \ee
\par
The system of brane equations (\ref{einstein}), due to the
presence of non-local bulk effects is not closed. Equations
(\ref{A}), (\ref{AB}) and (\ref{bianchi1}) consist a system of
three equations with four unknown functions $A, B, \mathcal{U},
\mathcal{P}$. To make these equations closed we will look for
solutions of the form $\underline{AB=1}$ and find the most general
braneworld metrics for such configurations. Then, equation
(\ref{AB}), outside the matter distribution is written
equivalently as \be L^{0}_{0}=L^{r}_{r}\,, \label{l00lrr} \ee and
equations (\ref{A}), (\ref{bianchi1}) get respectively the form
\be
\frac{A'}{A}=\frac{1-A}{r}+Ar\Big[\Lambda_{4}+\frac{3}{2}\alpha^{2}-\alpha
(2L^{0}_{0}+L^{\theta}_{\theta})\Big]\,, \label{A1} \ee \be
\frac{2}{r}(L^{0}_{0}-L^{\theta}_{\theta})+(2L^{0}_{0}+L^{\theta}_{\theta})\,'=0\,.
\label{bianchi2} \ee Therefore, we have to solve the above two
equations (\ref{A1}), (\ref{bianchi2}) with $L_{0}^{0},
\,L_{r}^{r},\, L_{\theta}^{\theta},\, L_{\phi}^{\phi}$ satisfying
(\ref{algebraic}), (\ref{lthth}), (\ref{l00lrr}) together with one
of the relations (\ref{case1}) or (\ref{case2}) or (\ref{case3}).
\par
To use relations (\ref{case1}), (\ref{case2}), (\ref{case3}) we
have first to identify $L_{1}^{1},\,L_{2}^{2},\,L_{3}^{3}$ with
$L_{r}^{r},\, L_{\theta}^{\theta}$ and $L_{\phi}^{\phi}$. In doing
so, we have to examine various cases depending on the choice of
$L$'s. Demanding to have non-vanishing
$\overline{\textsf{E}}^{\mu}_{\nu}$, only two cases remain, while
all the others have $\overline{\textsf{E}}^{\mu}_{\nu}=0$ and have
been thoroughly examined in \cite{pappa}. These two cases are \be
L^{r}_{r}=\sqrt{(L^{0}_{0})^{2}+\tau_{r}-\mathcal{T}^{0}_{0}}\,\,,\,\,
L^{\theta}_{\theta}=L^{\phi}_{\phi}=\epsilon
\sqrt{(L^{0}_{0})^{2}+\tau_{\theta}-\mathcal{T}^{0}_{0}}\,,
\label{cases} \ee depending on the sign $\epsilon =+1$ or $-1$. In
both cases, equation (\ref{l00lrr}) is equivalent to the relation
\be \tau_{r}=\mathcal{T}^{0}_{0} \Leftrightarrow
2\,\mathcal{U}+\mathcal{P}=0\,. \label{up} \ee This equation,
together with equation (\ref{algebraic}) which becomes \be
L^{0}_{0}\pm L^{0}_{0}+2\epsilon
\sqrt{(L^{0}_{0})^{2}+\mathcal{P}}=
\sqrt{4(L^{0}_{0})^{2}+4\Lambda_{4}-2\Lambda_{5}+3\alpha^{2}+2\mathcal{P}}\,,
\label{algebraic1} \ee relation
$L^{\theta}_{\theta}=L^{\phi}_{\phi}=\epsilon
\sqrt{(L^{0}_{0})^{2}+\mathcal{P}}$, and equations (\ref{A1}),
(\ref{bianchi2}) are everything we have to satisfy. It is obvious
that the two square roots appearing in (\ref{algebraic1}) have to
be well defined.

\section*{4. \,\,Black Hole Solutions including Non-Local Bulk Effects}
\subsection*{a. First solution}
\hspace{0.8cm} We first consider the $-$ of the $\pm$ sign in the
algebraic relation (\ref{algebraic1}). The only
 solution with non-zero  $\overline{\textsf{E}}^{\mu}_{\nu}$ corresponds
to
 $\epsilon = +1$ with $\mathcal{U}, \mathcal{P}$ being constants
\be
\mathcal{P}=-2\,\mathcal{U}=\frac{1}{2}\left(4\Lambda_{4}-2\Lambda_{5}+3\alpha^{2}\right)\,.
\label{UP} \ee Straightforward integration of (\ref{bianchi2})
gives \be
r=c\,\Big(\sqrt{(L^{0}_{0})^{2}+\mathcal{P}}+L^{0}_{0}\Big)^{\frac{1}{4}}\,\,
e^{\frac{3}{8\mathcal{P}}\left(\sqrt{(L^{0}_{0})^{2}+
\mathcal{P}}\,+\,L^{0}_{0}\right)^{2}}\,\,, \label{integrbianchi}
\ee with $c>0$ an integration constant. It is convenient to define
the positive dimensionless variable \be
z=\frac{9}{8|\mathcal{P}|}\Big(\sqrt{(L^{0}_{0})^{2}+\mathcal{P}}+L^{0}_{0}\Big)^{2}\,,
\label{zL} \ee and then, integration of the remaining equation
(\ref{A1}) gives
\begin{eqnarray}
B=\frac{1}{A}&=&1-\frac{\gamma}{r}-\beta r^{2} \nonumber\\
&&+\,\,sg(\zeta)\,\frac{\delta}{r}\,\Big[
\frac{128}{105}\,_{1}F_{1}\Big(\frac{15}{8},\frac{23}{8};sg(\zeta)z\Big)\,z+
\frac{9}{8}\Big(\frac{1}{z}-sg(\zeta)\frac{8}{7}\Big)
\,e\,^{sg(\zeta)\,z}\Big]z^{\frac{7}{8}}\,,
\label{integration}
\end{eqnarray}
where $\gamma$ is another integration constant (typically
interpreted as $2G_{N}M$ with $M$ being the mass of the point
particle, $G_{N}$ the Newton's constant), \be \beta =\frac{1}{3}
\Lambda_{4}+\frac{1}{2} \alpha^{2}
\,\,\,\,\,\,,\,\,\,\,\,\,\zeta=\frac{\alpha^2}{9}(4\Lambda_{4}-2\Lambda_{5}+3\alpha^{2})\,,
\label{beta} \ee and
$\delta=\frac{4}{9}(\frac{9}{8|\mathcal{P}|})^{\frac{1}{8}}\,|\mathcal{P}|
\alpha c^{3}>0$. Relation (\ref{integrbianchi}) becomes \be
r=\Big(\frac{\delta}{\sqrt{|\zeta|}}\Big)^{\frac{1}{3}}z^{\frac{1}{8}}\,e\,^{sg(\zeta)\,z/3}\,.
\label{laf} \ee The well-definiteness of the square roots in
(\ref{algebraic1}) and the fact that $L^{0}_{0} \geq 0$ (eqs.
(\ref{l00lrr}), (\ref{cases})) translate to $z\geq\frac{9}{8}$,
which means that $r$ is larger (smaller) than $(9 e^{3
sg(\zeta)}/8)^{1/8}\,(\delta / \sqrt{|\zeta|})^{1/3}$ for
$\zeta>0$ ($\zeta<0$). Note that $r(z)$ is a monotonically
increasing (decreasing) function for $\zeta > 0$ ($\zeta < 0$) in
its range of validity.
\par
\textit{Thus, the resulting solution is given in parametric form
by equations (\ref{integration}), (\ref{laf}), containing two
integration constants $\delta>0\,,\,\gamma$ and two parameters
$\beta\,,\, \zeta$ connected to $\alpha, \Lambda_{4}, \Lambda_{5}$
by relations (\ref{beta}). The electric components of the Weyl
tensor are given by equation (\ref{UP}),
$\mathcal{P}=-2\mathcal{U}=9\zeta/(2\alpha^2)$.}
\par
Comparing the solution (\ref{integration}) with the solution
(\ref{generalA>}) having $\overline{\textsf{E}}^{\mu}_{\nu}=0$, we
note that there is a new term, besides the conventional Newtonian
and $(A)dS_{4}$ terms, which carries the information of the
gravitational field in the bulk. For $\zeta > 0$, the asymptotic
behavior $r \rightarrow \infty$ of this new term in the solution
(\ref{integration}) is seen to be $AdS_{4}$\,--like, i.e.
$\sqrt{\zeta}\,r^2$. Thus, asymptotically, the effective
cosmological constant is $\beta-\sqrt{\zeta}$. For $\zeta < 0$,
the asymptotic behavior $r \rightarrow 0$ of the new term in the
solution (\ref{integration}) is Newtonian, i.e.
$-2\Gamma(7/8)\delta /r$. Thus, the effective Newton's constant in
this regime appears larger.
\begin{figure}[h!]
\centering
%\hspace{0.1cm}%
\begin{tabular}{cc}
\includegraphics*[width=200pt, height=150pt]{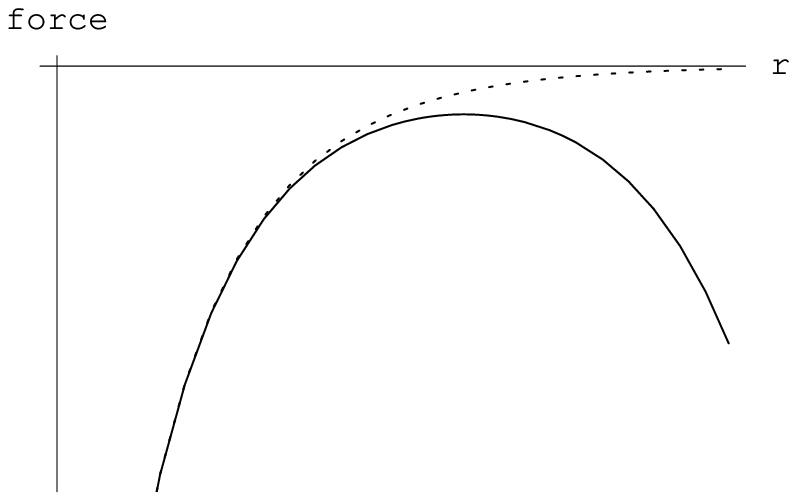}&%
%\hspace{0.1cm}%
\includegraphics*[width=200pt, height=150pt]{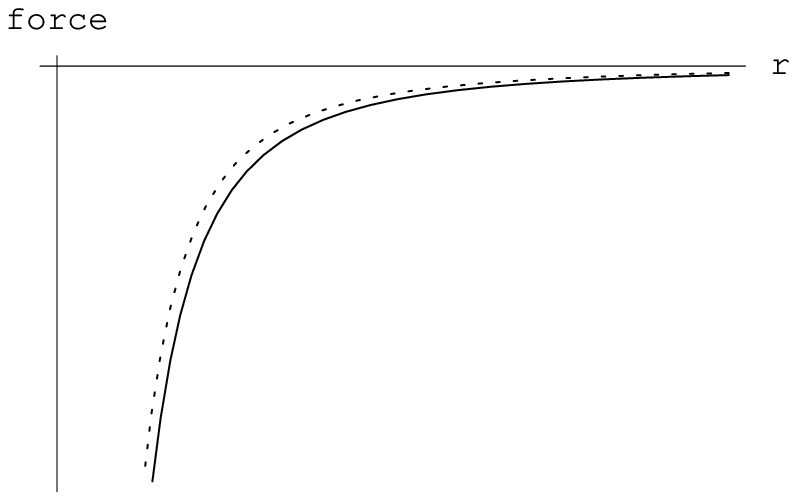}\\%
(a): $\zeta >0$, $\beta=0$&(b):  $\zeta <0$, $\beta=0$
\end{tabular}
 \caption{
Dotted lines represent the Newtonian force. Continuous lines
represent the total force, i.e. the sum of the Newtonian and the
new force.}
\end{figure}
\par
We know that for non-relativistic particles the effective
potential is $2\Phi=B-1$. As it can be seen, the new force
corresponding to the above non-local term is always attractive.
For $\zeta>0$, its magnitude is monotonically increasing with
distance, while for $\zeta<0$, this happens in decreasing
distances (after a characteristic scale). In order for the new
term not to disturb the well-measured Newtonian law at distances
from the $cm$ to the solar-distance scale, one has in both cases
to adjust the quantity $\delta / \gamma$ as small as desired. For
$\zeta>0$ and for larger distances, the sum of the Newtonian and
the new force decreases (in magnitude) slower than the Newtonian
force, while for even larger distances, this sum grows to infinity
(Figure 1a). For $\zeta<0$, deviations between the total and the
Newtonian force appear only at small distances (Figure 1b). The
$(A)dS_{4}$ term $\beta r^2$ is generally considered to be of
cosmological origin and is not considered here to be of importance
at the local level.
\par
Finishing with the above solution (\ref{integration}), we notice
that this may have some interesting physical implications. For
$\zeta>0$, because the total gravitational force grows slower than
the conventional Newtonian law, this force may serve as a possible
qualitative explanation for the yet unresolved problem of galactic
rotation curves. However, numerical fittings with real data remain
to be done. On the other hand, the solution with $\zeta<0$ could
be considered if submillimeter deviations from the Newtonian law
are discovered.
\subsection*{b. Second solution}
\hspace{0.8cm} Now, considering the $+$ case of the $\pm$ sign in
(\ref{algebraic1}), we define the dimensionless variable \be
v=2+\epsilon \sqrt{1+\frac{\mathcal{P}}{(L^{0}_{0})^2}}\,,
\label{v} \ee and equation (\ref{algebraic1}) gets the form \be
(L^{0}_{0})^{2}=\frac{4\,\zeta}{9\alpha^2\,(v^2-3)}\,, \label{la}
\ee where \be
\zeta=\frac{9\alpha^2}{8}\left(4\Lambda_{4}-2\Lambda_{5}+3\alpha^2\right).
\label{lagos} \ee Thus, the electric components of the Weyl tensor
are found in terms of $v$ to be \be
\mathcal{P}=-2\,\mathcal{U}=\frac{4\zeta\,(v-1)(v-3)}{9\alpha^2\,(v^{2}-3)}\,.
\label{tra} \ee The Bianchi equation (\ref{bianchi2}) can be seen
to become a simple separable differential equation for $v(r)$: \be
\frac{dv}{dr}+\frac{2}{3r}\,(v-3)(v^{2}-3)=0\,. \label{lop} \ee
Straightforward integration of (\ref{lop}) gives \be
r=\Big(\frac{\delta}{\sqrt{|\zeta|}}\Big)^{\frac{1}{3}}
\frac{|v-\sqrt{3}|^{\,(\sqrt{3}+1)/8}}{|v-3|^{1/4}\,\,|v+\sqrt{3}|^{\,(\sqrt{3}-1)/8}}\,\,,
\label{plo}  \ee with $\delta>0$ an integration constant.
\par For $\epsilon =+1$, it is $v>2$ (with $v \neq 3$) and from (\ref{la}) we have $\zeta>0$.
More specifically, for $2<v<3$ it is
$r>(2-\sqrt{3})^{\sqrt{3}/4}\,(\delta/\sqrt{|\zeta|})^{1/3}$ and
$r(v)$ is monotonically increasing, while for $v>3$ it is
$r>(\delta/\sqrt{|\zeta|})^{1/3}$ with $r(v)$ monotonically
decreasing. For $\epsilon =-1$, from (\ref{algebraic1}) it is
$\mathcal{P}<0$, thus (\ref{v}) gives $1<v<2$. Therefore, from
(\ref{la}) we have $1<v<\sqrt{3}$ for $\zeta<0$, and
$\sqrt{3}<v<2$ for $\zeta>0$. More specifically, for
$1<v<\sqrt{3}$ it is
$r<(\sqrt{3}-1)^{\sqrt{3}/4}\,(\delta/\sqrt{|\zeta|})^{1/3}/2^{(\sqrt{3}+1)/8}$
with $r(v)$ being monotonically decreasing, while for
$\sqrt{3}<v<2$ it is
$r<(2-\sqrt{3})^{\sqrt{3}/4}\,(\delta/\sqrt{|\zeta|})^{1/3}$ with
$r(v)$ monotonically increasing.
\par
The remaining Einstein equation (\ref{A1}) gives \be
B=\frac{1}{A}=1-\frac{\gamma}{r}-\beta r^{2}
\oplus\,\frac{\delta}{r}
\int{|v-\sqrt{3}|^{-\frac{3(3-\sqrt{3})}{8}}\,(v+\sqrt{3})^{-\frac{3(3+\sqrt{3})}{8}}
\,\frac{v}{|v-3|^{7/4}}\,\,dv}\,, \label{do} \ee where $\gamma$ is
another integration constant (typically interpreted as $2G_{N}M$
with $M$ being the mass of the point particle, $G_{N}$ the
Newton's constant) and \be
\beta=\frac{1}{3}\Lambda_{4}+\frac{1}{2}\alpha^{2}\,. \label{gr}
\ee The symbol $\oplus$ means $-$ for $\epsilon =+1, \,v>3$ or for
$\epsilon =-1, \,1<v<\sqrt{3}$\,; for $\epsilon =+1, \,2<v<3$ or
for $\epsilon =-1, \,\sqrt{3}<v<2$ it means $+$. The above
integral cannot be computed in terms of known functions. However,
this can be done in the asymptotic regimes $r \rightarrow \infty$
and $r \rightarrow 0$. For $r \rightarrow \infty$ ($\epsilon=+1$),
the new term in the solution (\ref{do}) becomes $AdS_{4}$ --like,
i.e. $(\sqrt{2\zeta}/3\sqrt{3})\,r^{2}$ and thus, asymptotically,
the effective cosmological constant is
$\beta-(\sqrt{2\zeta}/3\sqrt{3})$. For $r \rightarrow 0$
($\epsilon=-1$), approximating the integral in (\ref{do}) around
$v=\sqrt{3}$, we find that the new term scales as
$r^{2(2-\sqrt{3})}$\,, giving therefore extra attractive force
$1/r^{2\sqrt{3}-3}$\,. Numerical evaluation of the integral in
(\ref{do}) leads for $\epsilon=+1$ qualitatively to the same
picture as that of Figure 1a, where by adjusting the quantity
$\delta / \gamma$ as small as desired, deviations from Newtonian
law appear only at large distances. Similarly, for $\epsilon=-1$,
the picture for the solutions resembles qualitatively to that of
Figure 1b, where deviations from Newtonian law appear only at very
small distances.
\par
\textit{Thus, the resulting solution is given in parametric form
by equations (\ref{plo}), (\ref{do}), containing two integration
constants $\delta>0\,,\,\gamma$ and two parameters $\beta\,,\,
\zeta$ connected to $\alpha, \Lambda_{4}$, $\Lambda_{5}$ by
relations (\ref{lagos}), (\ref{gr}). The electric components of
the Weyl tensor are given by equation (\ref{tra}).}
\subsection*{c. Deflection of light}
\hspace{0.8cm} We have considered so far the motion of
non-relativistic particles. However, the motion of a freely
falling photon in a static isotropic gravitational field
(\ref{spherical}) is described \cite{weinberg} by the equation \be
\left(\frac{d\phi}{dr}\right)^{2}=\frac{A}{r^4}\left(\frac{1}{J^2
B}-\frac{1}{r^2}\right)^{-1}\,, \label{wein} \ee where $J$ is an
integration constant. In the cases where the solutions
(\ref{integration}), (\ref{do}) deviate from Newton's law at large
distances, it is seen from (\ref{wein}) that $d\phi/dr \rightarrow
0$ as $r \rightarrow \infty$, and thus, the photon moves in a
``straight'' line of the background geometry in that region (even
when a second horizon exists, we consider it of cosmological size
compared to the local distances of interest). More specifically,
at large distances, it arises from (\ref{wein}) that
$\phi(r)-\phi(\infty)\simeq
(\frac{1}{J^2}+\beta-\sqrt{\zeta})^{-1/2}\,\,\frac{1}{r}$ \,for
the solution (\ref{integration}), while for the solution
(\ref{do}), $\sqrt{\zeta}$ is replaced by
$\sqrt{2\zeta}/3\sqrt{3}$ in the last expression. This means that
the ``impact parameter'' $\textsf{b}$ is
$\textsf{b}=(\frac{1}{J^2}+\beta-\sqrt{\zeta})^{-1/2}$ for
(\ref{integration}) (and respectively for (\ref{do}) with the
change of $\sqrt{\zeta}$). For our solutions, the total deflection
angle in (\ref{wein}) cannot be computed explicitly. However, we
can understand the influence of the new term on the motion of a
photon and compare to the Newtonian deflection. For doing so, we
have to refer to two photons with the same ``initial conditions'',
i.e. the same impact parameter $\textsf{b}$, one moving in a
Schwarzschild-$(A)dS_{4}$ background (denoted by the subscript 1)
and the other in the background defined by the solutions
(\ref{integration}), (\ref{do}) (denoted by the subscript 2). The
following equations are easily obtained for the solutions
(\ref{integration}), (\ref{do}) respectively: \be
\frac{1}{\sqrt{\zeta}\,r^4}\frac{(dr_{1})^2-(dr_{2})^2}{(d\phi)^{2}}=\frac{\sqrt{z}}{e^{z}}
\left[\frac{128}{105}\,_{1}F_{1}\left(\frac{15}{8},\frac{23}{8};z\right)z+\frac{9}{8}
\left(\frac{1}{z}-\frac{8}{7}\right)e^{z}\right]-1\,,
\label{path1} \ee
\begin{eqnarray}
 &&\frac{3 \sqrt{3}}{\sqrt{2
\zeta}\,r^4}\frac{(dr_{1})^2-(dr_{2})^2}{(d \phi)^{2}}=\nonumber \\
&&\hspace{-8mm}\oplus
\frac{3\sqrt{3}}{\sqrt{2}}\frac{|v-3|^{\frac{3}{4}}\,(v+\sqrt{3})^{\frac{3(\sqrt{3}-1)}{8}}}{(v-\sqrt{3})
^{\frac{3(\sqrt{3}+1)}{8}}}
\int{(v-\sqrt{3})^{-\frac{3(3-\sqrt{3})}{8}}
(v+\sqrt{3})^{-\frac{3(3+\sqrt{3})}{8}}\frac{v
dv}{|v-3|^{7/4}}}-1\,. \label{path2}
\end{eqnarray}
It is obvious that for the branch $\epsilon=+1, v>3$ which extends
to infinity, the right-hand side of equation (\ref{path2}) is
negative, giving $(dr_{2})^2>(dr_{1})^2$. Therefore, extra
deflection of light compared to the Newtonian deflection arises.
This situation of increased deflection (compared to that caused
from the luminous matter) has been well observed in galaxies or
clusters of galaxies and the above solution might serve as a
possible way for providing an explanation. On the other hand, it
is easily checked that equation (\ref{path1}) provides less
deflection compared to Newtonian deflection at the distances of
interest.

%%%%%%%%%%%%%%%%%%%% END %%%%%%%%%%%% END %%%%%%%%%%%%%%%% END %%%%%%%%%%%%%%

%%%%%%%%%%%%%%%%%%%%%%%%%%% CONCLUSIONS %%%%%%%%%%%%%%%%%%%%%%%%%%%%%%%%%%%%%%%
\section*{5. \,\,\,Conclusions}
\hspace{0.8cm} In this paper, we presented a new class of brane
black hole solutions with induced gravity. It is known that the
non-local bulk effects, as they are expressed via the projection
of the Weyl tensor on the brane, do not make the brane dynamics
closed. We need to know the geometry of the bulk space in order to
be able to deal with the dynamics on the brane. In the case where
$g_{tt}=-g_{rr}^{-1}$, the system of equations consisting of the
modified Einstein equations and the Bianchi identities is closed
and we found all the possible black hole solutions. If we had to
look for more general spherically symmetric solutions, some
extra information would be needed for the non-local energy density
$\mathcal{U}$ or the non-local anisotropic stress $\mathcal{P}$.
\par
There has been argued \cite{ederys, edery} on kinematical grounds,
irrespectively from the gravitational dynamics, that the only
spherically symmetric geometries which may be candidates for explaining
from one side the extra deflection of light observed in
galaxies and clusters of galaxies and from the other side the
galactic rotation curves are of the form $g_{tt}=-g_{rr}^{-1}$.
However, severe criticism has appeared on this \cite{beke}. In the
present paper, we use this interesting and reasonable condition to
make the brane dynamics autonomous.
\par
The black hole solutions we found, are in a sense generalizations
of the spherically symmetric solutions we presented in
\cite{pappa}. They are representing strong-gravity corrections to
the spherically symmetric Schwarzschild-$(A)dS_{4}$ braneworld.
Their characteristic is that they predict a new
attractive force. There are classes of solutions with increasing $r$,
where this attractive force combined with the Newtonian one, results
to a net force which decreases slower than the Newton's force. This
might have interesting physical implications for the explanation
of galactic rotation curves. Within this class, a solution giving
extra deflection of light compared to General Relativity predictions
at galactic scales was found. It is interesting to observe that this solution
has non-trivial (i.e. not constant) non-local energy $\mathcal{U}$ and anisotropic
stress $\mathcal{P}$. In another class of solutions with
decreasing $r$, the new force starts to deviate from the Newton's
force at small distances, indicating that at submillimeter scale
we could have testable deviations from the Newtonian law.
\par
In our previous work we also had deviations from the Newton's law
at large distances. This deviation was caused by the presence of
the $(A)dS_{4}$ term $\beta r^2$, which for $\beta<0$ can also give extra
attraction. In our new solutions, the extra attractive force
appears because of the presence of a new term which only
asymptotically (when defined in this regime) behaves like $AdS_{4}$.
This new term arises because of the presence of the electric part
of the Weyl tensor and for an observer on the brane is a pure
non-local effect. We had also found in \cite{pappa} modifications
to Newton's law as a result of a change of the Newton's constant
due to the finite interior of the rigid object. This effect must
have an analogous contribution here if one solves the interior
problem.
\par
We have followed a braneworld viewpoint for obtaining braneworld
solutions, ignoring the exact bulk space. We have not provided a
description of the gravitational field in the bulk space, but
confined our interest to effects that can be measured by
brane-observers. By making assumptions for obtaining a closed
brane dynamics, there is no guarantee that the brane is embeddable
in a regular bulk. This is the case for a Friedmann brane
\cite{binetruy}, whose symmetries imply that the bulk is
Schwarzschild-$AdS_{5}$ \cite{muko, bow}. A Schwarzschild brane
can be embedded in a ``black string'' bulk metric, but this has
singularities \cite{hawking, gregory, giannakis, tamvakis,
wiseman}. The investigation of bulk backgrounds which reduce to
Schwarzschild-$(A)dS_{4}$ or more general black holes is in
progress.

%%%%%%%%%%%%%%%%%%%%%%%%%%%% BIBLIOGRAPHY %%%%%%%%%%%%%%%%%%%%%%%%%%%%%%%%%%%%%%

%%%%%%%%%%%%%%%%%%%%%%%%%%%%%%%%%%%%%%%%%%%%%%%%%%%%%%%%%%%%%%%%%%%%%%%%%%%%%%

 \end{document}